\documentclass[journal=jacsat,manuscript=article]{achemso}

\usepackage[version=3]{mhchem} 
\usepackage{ulem}
\usepackage{soul}
\usepackage{graphicx}
\newcommand{\angstrom}{\text{\normalfont\AA}}

\author{M. Saghir}
\affiliation{Department of Physics, University of Warwick, Coventry, CV4 7AL, United Kingdom}
\email{M.Saghir@warwick.ac.uk}
\author{A. M. Sanchez}
\affiliation{Department of Physics, University of Warwick, Coventry, CV4 7AL, United Kingdom}
\author{S. A. Hindmarsh}
\affiliation{Department of Physics, University of Warwick, Coventry, CV4 7AL, United Kingdom}
\author{S. J. York}
\affiliation{Department of Physics, University of Warwick, Coventry, CV4 7AL, United Kingdom}
\author{G. Balakrishnan}
\affiliation {Department of Physics, University of Warwick, Coventry, CV4 7AL, United Kingdom}
\email{G.Balakrishnan@warwick.ac.uk}

\title{Nanomaterials of the topological crystalline insulators, Pb$_{1-x}$Sn$_x$Te and Pb$_{1-x}$Sn$_x$Se.}

\abbreviations{TI,TCI,ARPES,EDAX,EBSD,SEM,VLS,XRD}

\begin{document}
\normalem
\begin{abstract}
Topological insulators (TIs) and topological crystal insulators (TCIs) exhibit exotic surface properties. We present optimised growth procedures to obtain high quality bulk crystals of the TCIs Pb$_{1-x}$Sn$_x$Te and Pb$_{1-x}$Sn$_x$Se, and nanowires from the bulk crystals using the vapour-liquid-solid (VLS) technique. Nanowires of Pb$_{1-x}$Sn$_x$Te have been produced with a Sn composition of $\sim$ $x = 0.25$, at which a transition from trivial to non-trivial insulator is reported. The results obtained on the growth of nanomaterials of Pb$_{1-x}$Sn$_x$Se are also described. Detailed characterisation of the bulk crystals and the nanomaterials through x-ray diffraction, microscopy techniques and EDX analysis are presented.
\end{abstract}
\maketitle

In recent years topological insulators (TIs) have been the subject of much interest. Materials such as HgTe, Bi$_2$Se$_3$ and Bi$_2$Te$_3$ are some of the first compounds that were confirmed experimentally to exhibit metallic gapless surface states, typical of TI's. Prompted by the discovery of the 2D and 3D TIs, there has since been a focus to discover new materials with topologically protected states. This has led to the discovery of the topological crystalline insulators (TCIs), which are a subclass of TIs. TCIs differ from TIs in that the band degeneracy observed is protected by mirror plane symmetry replacing the role played by time-reversal symmetry in conventional TIs.

SnTe was the first material, proposed by Hsieh \emph{et al.}\cite{Hsieh2012}, to exhibit TCI properties. Ab-initio calculations show that band inversions occur at four L-points in the 3D Brillioun zone - a typical signature for the electronic structure of TCIs. This was later experimentally determined on the (100) surface of SnTe, by angle-resolved photoelectron spectroscopy (ARPES).\cite{Hsieh2012} 

The effect of substituting Sn in PbTe and PbSe has been explored and it was suggested that by introducing strain into the lattice of these materials, a band inversion could be induced, thereby changing the trivial insulating nature of the materials to topologically non-trivial.\cite{Fu2007} As a result, in the solid solutions Pb$_{1-x}$Sn$_x$Te and Pb$_{1-x}$Sn$_x$Se, TCI surface states have been experimentally observed.\cite{Xu2012,Dziawa2012,Tanaka2013} The TCI nature of these materials is unaffected by the mixing disorder of the system.\cite{Safaei2013} Further experimental evidence obtained using spin-resolved photoelectron spectroscopy (SRPES) has led to the observation of spin textures for the (001) metallic surfaces in both Pb$_{0.73}$Sn$_{0.27}$Se and Pb$_{0.60}$Sn$_{0.40}$Te.\cite{Wojek2013,Xu2012}

Unlike in the case of SnTe, where the experimental observation of the TCI states can be difficult due to the p-type nature of the material, the tunable nature of the chemical potential to n and p-type in Pb$_{1-x}$Sn$_{x}$Te/Se makes these compounds more suitable for experimentally observing the TCI states, thus providing the motivation for the study of these materials.\cite{Dziawa2012}

The resulting solid solution upon substitution, Pb$_{1-x}$Sn$_x$Te, is a narrow band semiconductor with a tunable electronic structure based on the Sn/Pb ratio.\cite{Xu2012} The structure of Pb$_{1-x}$Sn$_x$Te is cubic for all Sn substitutions. For Sn substitution of up to $x$ $\le$ 0.4 in Pb$_{1-x}$Sn$_x$Se, the structure remains cubic, similar to that of PbSe, whilst the other end member SnSe adopts an orthorhombic structure. 

For both Pb$_{1-x}$Sn$_x$Te and Pb$_{1-x}$Sn$_x$Se, the location of the band inversions are analogous to the material SnTe. The introduction of Sn atoms allows for the closing of the band gap producing a Dirac state. For Pb$_{1-x}$Sn$_x$Te, the onset for the critical transition from a trivial to non-trivial insulator, occurs at a Sn substitution of $\sim$ $x = 0.25$ and the optimum point of this transition is $x = 0.4$.\cite{Tanaka2013,Xu2012}. A TCI phase onset is observed in Pb$_{1-x}$Sn$_{x}$Se for $x$ values between $0.18$ $\le$ $x$ $\le$ $0.3$. This transition is further dependent on a critical temperature, up to T$_c$ = 250 K~\cite{Dziawa2012} and the TCI transition is not observed above this critical temperature.  

It has been difficult to investigate the surface properties of TIs and TCIs, as the bulk signal all too often suppresses the signal from the protected surface state. This difficulty can be addressed by growing nanostructures, thus increasing the surface-area-to-volume (SAVR) of these materials, to better observe the metallic surface states.\cite{Kong2010a, Cha2010} The growth of the TIs, Bi$_{2}$Se$_{3}$/Te$_{3}$, in 2D and nanoform, has been well investigated and reported.\cite{Kong2010b,Wang2013,Li2011} Nanomaterials of the first TCI, SnTe, has also been reported by us in our previous work and by others.\cite{Saghir2014, Li2013, Safdar2013} The nanomaterial growth was achieved using a vapour-liquid-solid (VLS) growth technique as this is a proven technique for obtaining high quality nanomaterials. \cite{Lee2008, Lieber2011, Yang2005, Gao2002,Medlin2010}.

We have chosen to study the formation of nanomaterials starting with bulk crystals of Pb$_{1-x}$Sn$_x$Te, for an optimum Sn substitution level of $x$ $= 0.4$ and Pb$_{1-x}$Sn$_x$Se with three different Sn substitution levels $x$ $=0.18, 0.23$ and $0.30$. We report the experimental evidence for the growth of high quality single crystal nanowires of Pb$_{1-x}$Sn$_x$Te ($x$ $= 0.23(2)$), close to the critical Sn content at which the TCI transition occurs in this material. We also report attempts to produce nanomaterials of Pb$_{1-x}$Sn$_x$Se. In both cases, a Au-catalysed VLS growth technique was used, similar to that adopted for SnTe\cite{Saghir2014}. The optimum growth conditions and parameters obtained from the study are presented. The methods described herein provide a route to producing suitable nanomaterials which may lead to the successful investigation of the enhanced TCI states of the compounds Pb$_{1-x}$Sn$_{x}$Te/Se. Detailed characterisation performed on the bulk crystals grown, as precursors to the nanomaterial growth, include powder x-ray diffraction (XRD), x-ray Laue diffraction, scanning electron microscopy (SEM) and energy dispersive x-ray analysis (EDX).  The resultant nano and micro materials obtained are characterised using SEM, EDX, transmission electron microscopy (TEM) and selective area electron diffraction (SAED).

Starting with high purity Pb and Sn shot (Alpha Aesar 99.99 $\%$) and powders of Sn and Se, (Alpha Aesar 99.99 $\%$), the stoichiometirc mixtures were carefully placed in quartz tubes. The tubes were evacuated and sealed under vacuum. Following the procedure described by Tanaka \emph{et al.}, a modified Bridgman was employed. The quartz tubes were placed vertically into a furnace and the temperature was ramped to $\approx$ 1075 $^{\circ}\mathrm{C}$ and remained at this temperature for a period of 48 hours.{\cite{Strauss1967} The furnace was then slow cooled at 2 $^{\circ}\mathrm{C}$/h to 700 $^{\circ}\mathrm{C}$. The tubes were rapidly cooled from this temperature to room temperature at a rate of 200 $^{\circ}\mathrm{C}$/h.

The crystal boules obtained had shiny metallic surfaces when examined visually. The phase purity of the crystals was determined by performing XRD. Small pieces of the crystal boules were finely powdered and placed in a Panalytical X'Pert Pro system with a monochromatic Cu K$\alpha$1 radiation source. Figure \ref{fig:XRD} shows the X-ray diffraction data obtained for two samples, Pb$_{0.60}$Sn$_{0.40}$Te and Pb$_{0.70}$Sn$_{0.30}$Se. The x-ray patterns obtained suggests that the crystals grown are indeed single phase with lattice parameters in good agreement with the expected values (see Table \ref{tab:EDAX}). 

\begin{figure}[tb]
\centering
\includegraphics[width=0.8\columnwidth]{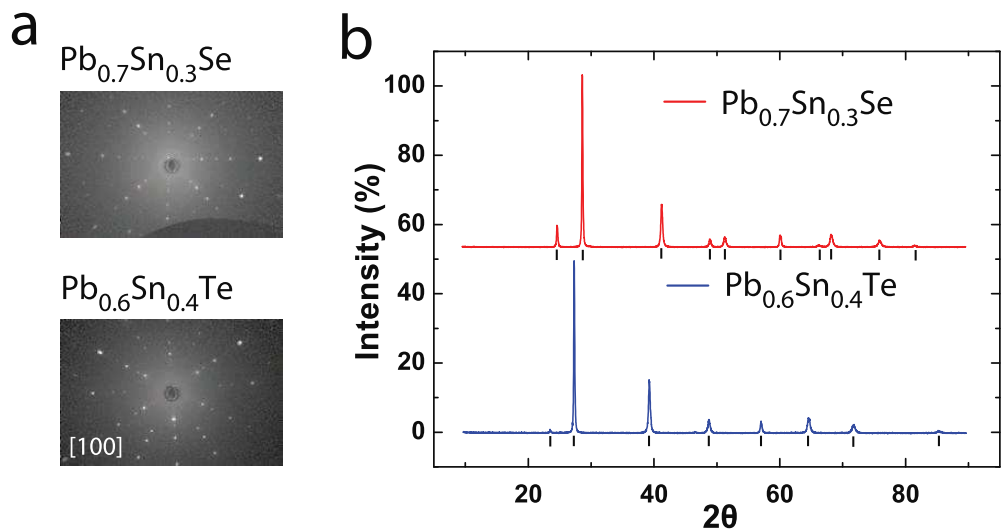}
\caption{(a) X-ray Laue diffraction data taken of the crystals along the [100] direction. The sharp spots demonstrate the high crystallinity of samples. (b) Powder XRD spectra taken on crushed powders from the as grown boules of Pb$_{0.70}$Sn$_{0.30}$Se and Pb$_{0.60}$Sn$_{0.40}$Te demonstrating the single phase nature of the crystal boules. The tick marks show the positions of the expected Bragg peaks.}
\label{fig:XRD}
\end{figure}

The crystallinity of the boules produced were examined using X-Ray Laue diffraction. This was performed across the surface of the crystal boules at various points. The diffraction patterns observed revealed sharp spots as can be seen in Figure \ref{fig:XRD}a. This demonstrates the high crystalline nature of the samples grown and provides the orientation of the crystals cleaved from the as-grown boules.

Compositional analysis was performed on different sections of the crystal boules. Cleaved sections of the boules were examined using an EDX system on a Zeiss SUPRA 55-VP scanning electron microscope. The results showed that all the crystals grown were of a stoichiometry similar to the nominal starting compositions, within error, as shown in Table \ref{tab:EDAX}. The crystals were powdered for use as starting materials for the growth of nanomaterials.

\begin{table*}
\footnotesize
\caption{Representative atomic compositions of the bulk Pb$_{0.60}$Sn$_{0.40}$Te and Pb$_{1-x}$Sn$_{x}$Se crystal boules are shown below. The data was obtained using EDX analysis of the bulk crystals. Lattice parameters for the powdered sections of the crystal boules obtained using powder XRD are also presented. The lattice parameters obtained show a slight discrepancy when compared to previously published data.\cite{Strauss1967} This discrepancy may be possible due to the strain in the samples. Further studies are required to investigate this.} 
\centering 
\begin{tabular}{c|c c c c |c } 
\hline 
Nominal starting composition & \multicolumn{4}{c|}{Atomic Percent (\%)} &  Lattice  \\
 & Pb & Sn & Te/Se & Total & parameter, \emph{a} (\AA) \\ 
\hline 
Pb$_{0.60}$Sn$_{0.40}$Te & 29(2) & 21(2) & 50(2) & 100 & 6.420(0.020) \\  
Pb$_{0.82}$Sn$_{0.18}$Se & 41(2) & 9(2) & 50(2) & 100 & 6.147(0.004) \\  
Pb$_{0.77}$Sn$_{0.23}$Se & 40(2) & 11(2) & 48(2) & 100 & 6.147(0.005) \\  
Pb$_{0.07}$Sn$_{0.30}$Se & 34(2) & 16(2) & 49(2) & 100 & 6.146(0.008) \\  
\hline 
\hline 
\end{tabular} 
\label{tab:EDAX} 
\end{table*}

Various parameters were refined and optimised to find the best conditions to obtain high quality nanowires with a good yield. This involved performing over 20 experiments adjusting the temperature of the hot zone as well as substrate position and argon flow rate similar to methods adopted in our previous work on the TCI, SnTe.\cite{Saghir2014}

The preparation of the silicon substrates used for nanomaterial growth involved a two step process. First the silicon wafer substrates (approx. 5 mm x 30 mm) were cleaned using a 50:50 mixture of acetone and isopropan-2-ol. These were then allowed to dry naturally in air before suspending a sodium citrate gold nanoparticle buffer solution on the surface of the substrates (Alpha Aesar 20 nm gold nanoparticles). The buffer was held in position by the surface tension of the solution on the substrates. The solution was then allowed to evaporate in ambient room temperature conditions. We found that the Au nanoparticles dispersed on the surface with a density of $\approx$ 5/$\mu$m$^2$.

\begin{figure}[tb]
\centering
\includegraphics[width=0.8\columnwidth]{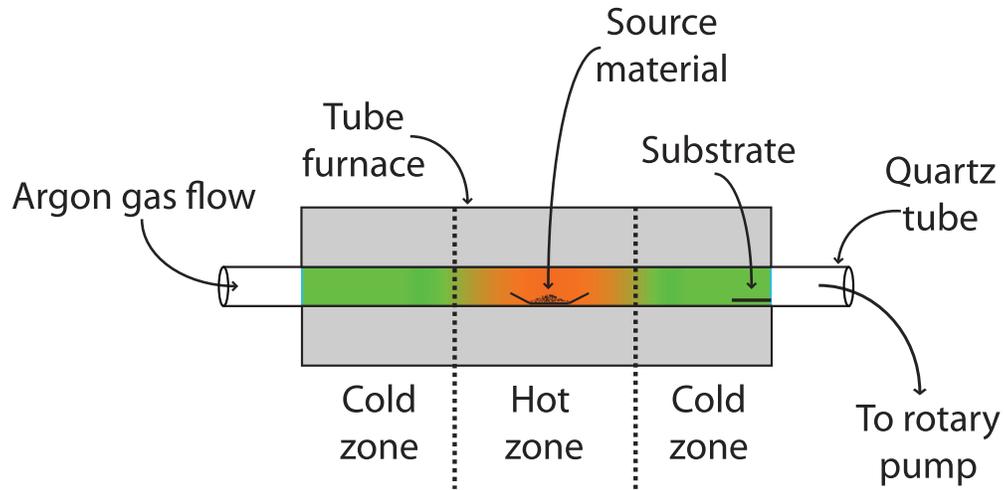}
\caption{Schematic of the quartz tube furnace used for the nanomaterial growths whilst also showing the positions of the source material and substrate.}
\label{fig:Schematic}
\end{figure}

5 mg of the source material was then placed at the centre of an alumina-silicate boat which in turn was placed in the centre of the `hot zone' of a 20 cm tube furnace. The silicon substrates were placed down stream in a `cold zone'. For the growth of Pb$_{0.80}$Sn$_{0.20}$Te nanowires, the furnace was rapidly heated to 540 $^{\circ}\mathrm{C}$ in 25 minutes under a flow of argon, to suppress oxide growth and to act as a carrier gas for the source material. The furnace remained at this temperature for a further 120 minutes at which point it was allowed to cool to room temperature naturally. The temperature of the `cold zone' where substrates were placed for micro and nanomaterial growth was $\approx$ $300$ $^{\circ}\mathrm{C}$. The furnace used with the source and substrate arrangements are shown in Figure \ref{fig:Schematic}.

After performing the nanomaterial growth, the substrates were removed from the furnace and upon visual inspection, metallic grey features on the surface of the substrates could be seen. The inside of the quartz tube was found to be coated with a thin metallic layer at the `cold zone'. Table \ref{tab:Results} shows a summary of the growth results obtained for the various starting materials. For the growths performed with Pb$_{1-x}$Sn$_{x}$Te, nanomaterials in the form of wires were obtained. The growth of microcrystals surrounding the free standing nanowires was also observed. For the growths with Pb$_{1-x}$Sn$_{x}$Se, however, predominantly microcrystals in the form of cubes were observed with a few nanowires present.

\begin{table*}
\footnotesize
\caption{Summary of nanomaterial growth results and chemical composition of the materials obtained using EDX analysis.} 
\centering 
\begin{tabular}{c| c| c| c| c} 
\hline 
Starting material (Powder) & Nanowires & Composition & Microcrystals & Composition \\ [0.5ex] 
\hline 
Pb$_{0.60}$Sn$_{0.40}$Te & Yes & Pb$_{0.77(2)}$Sn$_{0.23(2)}$Te & Yes & Pb$_{0.60(2)}$Sn$_{0.40(2)}$Te \\ 
Pb$_{0.82}$Sn$_{0.18}$Se & No & - & No & -  \\ 
Pb$_{0.77}$Sn$_{0.23}$Se & No & - & No & -  \\ 
Pb$_{0.70}$Sn$_{0.30}$Se & Yes & SnSe & Yes & PbSe \\  
\hline 
\hline 
\end{tabular} 
\label{tab:Results} 
\end{table*}

TEM was used to obtain information on the quality of the Pb$_{1-x}$Sn$_{x}$Te nanowires grown which had a composition of Pb$_{0.77(2}$Sn$_{0.23(2)}$Te when examined by EDX in TEM mode as shown in Table \ref{tab:Results}. The nanowires were found to be between 10 and 50 $\mu$m long with a typical thickness of $\approx$ 100 nm. The growth density of the nanowires, observed from SEM, was found to be $\approx$ 0.40/$\mu$m$^2$. Figure \ref{fig:SEM} shows an SEM image of the results for a typical growth with both nanowires and microcrystals of Pb$_{1-x}$Sn$_{x}$Te obtained.

\begin{figure}[tb]
\centering
\includegraphics[width=0.34\columnwidth]{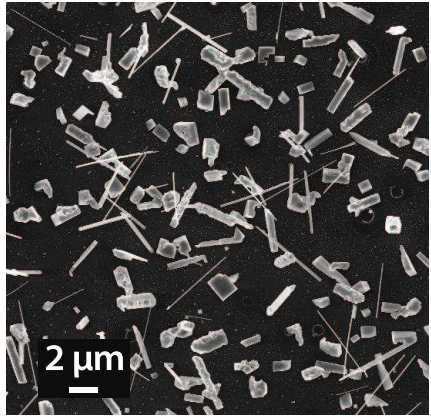}
\caption{Representative SEM image of the growth of Pb$_{1-x}$Sn$_{x}$Te nanowires and microcrystals. The thickness of the nanowires are $\approx$ 100 nm and lengths of up to 15 $\mu$m are observed. The microcrystals are distributed randomly but have distinct cubic growth facets.}
\label{fig:SEM}
\end{figure}

It has been widely reported that dispersed gold nanoparticles on the surface of the silicon substrates act as catalysts and promote the growth of nanowires.\cite{Kirkham2007} This is achieved by forming an alloy at the tip of the nanowire. The TEM image (Figure \ref{fig:TEM}a) shows a nanowire with a gold alloy which has formed at the tip, commonly seen in shorter ($<$ 4 $\mu$m) thinner nanowires ($<$ 80 nm). By observing various stages of the growth from nucleation, we observe that the alloy travels upwards in the growth direction of the nanowires. This is the typical tip-growth mechanism by which these nanowires grow.

Using TEM and SAED, the growth orientation of the crystalline nanowires was determined. Figure \ref{fig:TEM}b shows an atomic resolution image of a Pb$_{0.77(2)}$Sn$_{0.23(2)}$Te nanowire from which the lattice parameter was obtained (6.497(3) $\angstrom$). Using a focused ion beam (FIB) to prepare a nanowire in cross-section, SAED was performed to obtain structural information about the growth orientation of the nanowires (Figure \ref{fig:TEM}c). The nanowires were found to grow in the [100] direction. Sharp diffraction spots are also indicative of the highly crystalline nature of the nanowires.

We also determined from TEM that some thicker nanowires ($>$ 80 nm) grew with a core-shell structure as commonly observed in GaAs and ZnO nanowires (Figure \ref{fig:TEM}d).\cite{Kirkham2007,Ringer2000} When performing cross-sectional TEM on samples prepared using FIB milling, EDX does not show a variation in composition of the core-shell to that of the surface. We believe therefore that a transition in the growth process occurs, namely, from a VLS to VS growth process similar to that observed for AlGaAs nanowires.\cite{Wu2004} The axial growth of the central core is thought to be a VLS process after which a radial VS growth process dominates. The central core has a thickness comparable to the gold alloys at the tip of the nanowires which is approximately 30 nm.

Within the nanowires, we also observe contrast that agrees with Guinier-Preston like zones (G-P zones) under TEM (Figure \ref{fig:TEM}e). This metallurgical process is believed to occur at room temperature and is typically observed in age hardened aluminium alloys.\cite{Ringer2000} We find no structural defects such as dislocations or stacking faults within the regions and the G-P like zones appear at right angles to each other in line with the crystal lattice. EDX was unable to detect any compositional difference of the G-P zones to the surrounding matrix. A compositional difference is something that would be expected for G-P zones, however, the zones are surrounded by a thick matrix and therefore a compositional difference cannot be detected. As a result, the origin of these zones remains to be ascertained.

\begin{figure}[tb]
\centering
\includegraphics[width=0.4\columnwidth]{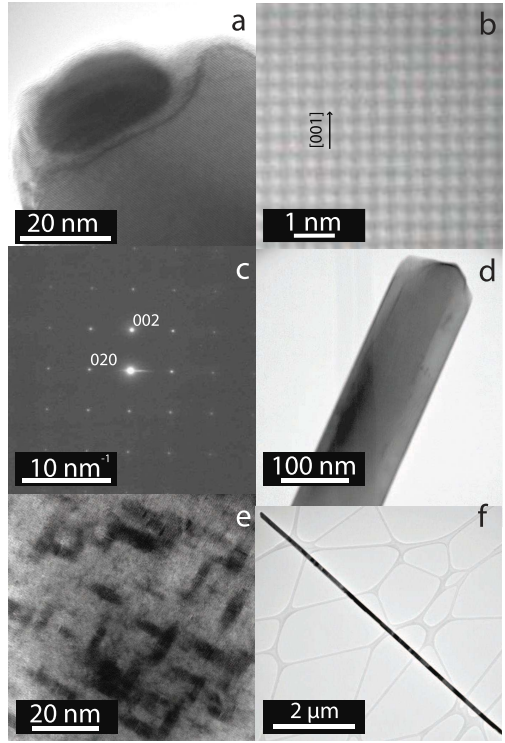}
\caption{(a) HR-TEM of the gold alloy formed at the tip of a Pb$_{0.77(2)}$Sn$_{0.23(2)}$Te nanowire which has been isolated on a TEM grid with carbon lace. (b) HR-TEM of a Pb$_{0.77(2)}$Sn$_{0.23(2)}$Te nanowire. A regular lattice can be seen showing the high crystalline nature of the the structure where the lattice parameter equates to $6.497(3)$ $\angstrom$. (c) Core-shell growth of nanowire. (d) SAED of nanowire. (e) Guinier-Preston like zones can be seen forming in-plane within the 3D lattice. No defects can be seen in these regions and compositional analysis reveals the they are the same composition as surrounding areas. (f) A typical long nanowire ($>$ 4 $\mu$m).}
\label{fig:TEM}
\end{figure}

For very long ($>$ 4 $\mu$m) and thick nanowires ($>$ 80 nm) as seen in figure \ref{fig:TEM}f, no gold alloy was found at the tip of the nanowire. From EDX, small trace amounts of gold could be detected which may suggest that the gold nanoparticle eventually becomes consumed within the body of the longer nanowire. In shorter nanowires ($<$ 4 $\mu$m), we observe a distinct gold alloy formation at the tip of the nanowire (\ref{fig:TEM}a). This further demonstrates the change in the growth mechanism of the nanowires - namely from an initial VLS process to form shorter thinner nanowires, to a VS process that leads to longer and thicker wires.

EDX obtained in TEM mode reveals that the nanowires have a chemical composition, within experimental error, that lies in the region of the critical transition point in Pb$_{1-x}$Sn$_{x}$Te nanowires ($\sim$ $x = 0.25$)\cite{Tanaka2013}, the point at which the material changes from a trivial insulator to a TCI (see Table \ref{tab:EDAX2}).
To compare the nanowires with the bulk crystals grown, TEM was used to obtain the lattice parameter from the edge of the nanowire. The lattice parameter matches (within experimental error) that of the starting powder obtained from crushed crystals shown in Table \ref{tab:EDAX}.

In addition to the nanowires, micron sized crystals were also observed in areas surrounding the nanowires. EBSD allows us to ascertain the orientation of the faces of the microcrystals grown. EDX compositional analysis of these microcrystals revealed a stoichiometry similar to that of the source material. Futhermore, no gold was present in the microcrystals suggesting a vapour-solid growth mechanism for these. Figure \ref{fig:SEM3}a shows the typical cubic Pb$_{0.60(2)}$Sn$_{0.40(2)}$Te microcrystals obtained in the growth. EBSD data obtained for the faces of the cubes showed they had a predominantly vicinal $\left\langle001\right\rangle$ orientation.

The distribution density (DD) of the microcrystals across the substrate surface was much greater than that for nanowires. We found that the DD increased towards the hotter end of the substrate, where we also saw fewer nanowires growing. Figure \ref{fig:SEM3}b also shows the typical structures observed as you move closer to the hotter end ($\approx$ 300 $^{\circ}\mathrm{C}$) of the substrate. The nucleation of the microcrystals increases greatly and they are in some cases approximately three times larger in size. As a result they merge forming a layer that is $\approx$ 10 $\mu$m thick, under which no substrate is visible. Conversely, for growths in the furnace at temperatures lower than the optimum (below $\approx$ 520 $^{\circ}\mathrm{C}$), very little nucleation is observed. Here only a few microcrystals are seen and the DD decreases by approximately one order of magnitude.

\begin{figure}[tb]
\centering
\includegraphics[width=0.5\columnwidth]{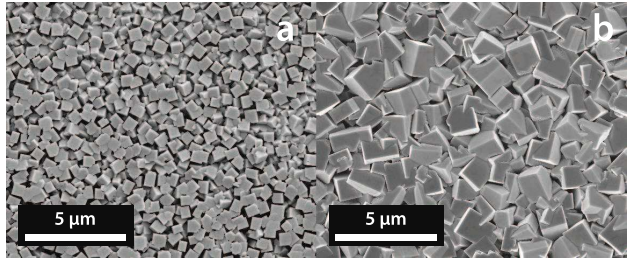}
\caption{(a) SEM image of Pb$_{0.60}$Sn$_{0.40}$Te microcrystals. (b) Larger microcrystals merge to form a thick layer of growth towards the hotter end of the substrate}
\label{fig:SEM3}
\end{figure}

Similar growth conditions used for the formation of Pb$_{0.77(2)}$Sn$_{0.23(2)}$Te nanowires were initially used for the compound Pb$_{1-x}$Sn$_{x}$Se. Under these conditions, the furnace temperature in the hot zone is $\approx$ 540 $^{\circ}\mathrm{C}$. At this temperature, and for all Sn compositions, we did not observe the growth of any nanowires or microcrystals. Instead, a thin layer of PbSe material was found to be deposited with trace amounts of Sn detected.

When the furnace temperature was increased to 550 $^{\circ}\mathrm{C}$, we did not observe any nanowire or microcrystal growth for Sn compositions $x = 0.18$ \& $x = 0.23$. However, starting with a Sn composition of $x = 0.30$, we found that the Pb$_{0.70}$Sn$_{0.30}$Se powder decomposed into PbSe microcrystals in the form of cubes and SnSe in the form of zig-zag nanowires (Figure \ref{fig:SEM2}). For temperatures higher than 550 $^{\circ}\mathrm{C}$, no nanowires were observed and the nucleation density of microcubes had increased to the point where the cubes merged to form a thick continuous layer, similar to the morphology seen in Pb$_{0.60(2)}$Sn$_{0.40(2)}$Te microcrystals (Fig.~\ref{fig:SEM3}b).

\begin{figure}[tb]
\centering
\includegraphics[width=0.5\columnwidth]{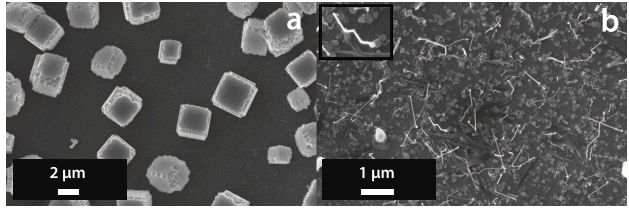}
\caption{(a) Representative SEM image of the growth of PbSe microcubes. (b) SnSe zig-zag nanowires with the inset for clarity. The thickness of the nanowires obtained for Pb$_{0.70}$Sn$_{0.30}$Se are $\approx$ 40 nm.}
\label{fig:SEM2}
\end{figure}

The microcubes were $\approx$ 1 $\mu$m$^3$ and the zig-zag nanowires were found to be $\approx$ 1 $\mu$m in length and $\approx$ 20 nm thick. From EBSD, the structure of the PbSe microcubes was found to be analagous to those of the Pb$_{0.60}$Sn$_{0.40}$Te microcrystals. Figure \ref{fig:EBSD} shows EBSD data where some features have been highlighted for ease of reference. It shows a typical microcube with growth faces in the $\left\langle001\right\rangle$ orientation. We did not observe any epitaxial relationship between the substrate and any of the various growth morphologies obtained for solid solutions of Pb$_{1-x}$Sn$_x$Te and Pb$_{1-x}$Sn$_x$Se.

\begin{figure}[tb]
\centering
\includegraphics[width=0.5\columnwidth]{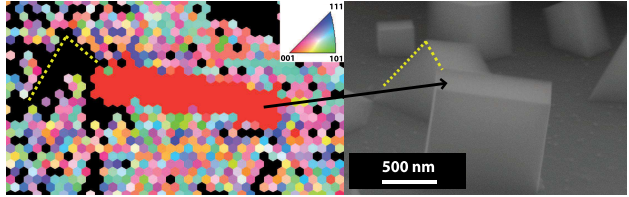}
\caption{Left: EBSD showing the top face of a typical PbSe microcube. The red region indicates a $\left\langle001\right\rangle$ face. The dashed yellow line is for reference and black arrow indicates the face of the microcube examined using EBSD.}
\label{fig:EBSD}
\end{figure}

\begin{table}
\footnotesize
\caption{Representative atomic compositions of the PbSe microcubes and SnSe zig-zag nanowires obtained using EDX analysis.} 
\centering 
\begin{tabular}{c| c c c| c} 
\hline  
Nominal starting composition & \multicolumn{3}{c|}{Atomic Percent (\%)} &  \\ 
Pb$_{0.70}$Sn$_{0.30}$Se & Pb & Se & Sn & Total \\  
\hline  
SnSe zig-zag nanowires & - & 50(2) & 50(2) & 100  \\  
PbSe microcubes & 50(2) & 50(2) & - & 100  \\ [1ex] 
\hline 
\hline 
\end{tabular} 
\label{tab:EDAX2} 
\end{table}

In summary we have outlined the optimal methods for the growth of high quality crystal boules of the TCIs Pb$_{0.60}$Sn$_{0.40}$Te and Pb$_{1-x}$Sn$_x$Se ($x = 0.18, 0.23$ \& $0.30$). From these we have demonstrated evidence for the growth of high quality Pb$_{0.77(2)}$Sn$_{0.23(2)}$Te nanowires using the VLS growth technique. The composition of the nanowires obtained for Pb$_{1-x}$Sn$_x$Te is similar to that where a TCI transition is known to occur in bulk crystals. The growth orientation of the Pb$_{0.77(2)}$Sn$_{0.23(2)}$Te nanowires was found to be in the [100] direction. It was found that using gold nanoparticles to activate the growth of nanowires is essential. We also find that there is a shift away from a VLS process to a VS process for very long nanowires ($>$ 4 $\mu$m) and that the nanowires grow with a core-shell model.
For the solid solution Pb$_{0.70}$Sn$_{0.30}$Se, we found the compound decomposed to form PbSe microcubes and SnSe zig-zag nanowires. For Pb$_{0.82}$Sn$_{0.18}$Se and Pb$_{0.77}$Sn$_{0.23}$Se, only a thin layer of PbSe material was deposited with trace amounts of Sn.
Our findings demonstrate reliable methods to make TCIs with high SAVR, which could be used to investigate enhanced TCI features. These nanomaterials are therefore good candidates to study TCI materials with enhanced surface effects and may be exploited for new applications.

\begin{acknowledgement}
This work was supported by the EPSRC, UK (EP/L014963/1). Some of the equipment used in this research was obtained through the Science City Advanced Materials Project, Creating and Characterizing Next Generation Advanced Materials Project, with support from AdvantageWest Midlands (AWM), and was partially funded by the European Regional Development Fund (ERDF). The authors thank T. Green, N. Sizer and A. Marsden for help with some of the work. We also thank T. E. Orton for valuable technical support.
\end{acknowledgement}

\end{document}